\documentclass[conference]{IEEEtran}
\usepackage{graphicx}
\usepackage{epstopdf}
\usepackage[latin1]{inputenc}
\usepackage{amsmath,amssymb,amscd,latexsym,dsfont}
\usepackage[english]{babel}
\usepackage{tabularx,cite}
\usepackage{graphicx}
\usepackage{algpseudocode}
\usepackage{algorithm}
\usepackage{algorithmicx}
\usepackage{float}
\usepackage{multicol}
\usepackage{psfrag}
\usepackage{comment,psfrag,subfigure,enumerate}
\usepackage{color}
\usepackage{amsthm}

\ifCLASSINFOpdf
\else
\fi
\begin{document}
%
\title{Green communication via Type-I ARQ: Finite block-length analysis}

\author{
    \IEEEauthorblockN{Behrooz Makki\IEEEauthorrefmark{1}, Tommy Svensson\IEEEauthorrefmark{1}, Michele Zorzi\IEEEauthorrefmark{2}}
    \IEEEauthorblockA{\IEEEauthorrefmark{1}Department of Signals and Systems, Chalmers University of Technology, Gothenburg, Sweden
    \\\{behrooz.makki, tommy.svensson\}@chalmers.se}
    \IEEEauthorblockA{\IEEEauthorrefmark{2}Department of Information Engineering, University of Padova, Padova, Italy
    \\zorzi@ing.unife.it}
}
\maketitle

\begin{abstract}
This paper studies the effect of optimal power allocation on the performance of communication systems utilizing automatic repeat request (ARQ). Considering Type-I ARQ, the problem is cast as the minimization of the outage probability subject to an average power constraint. The analysis is based on some recent results on the achievable rates of finite-length codes and we investigate the effect of codewords length on the performance of ARQ-based systems. We show that the performance of ARQ protocols is (almost) insensitive to the length of the codewords, for codewords of length $\ge 50$ channel uses. Also, optimal power allocation improves the power efficiency of the ARQ-based systems substantially. For instance, consider a Rayleigh fading channel, codewords of rate $1$ nats-per-channel-use and outage probability $10^{-3}.$ Then, with a maximum of 2 and 3 transmissions, the implementation of power-adaptive ARQ reduces the average power, compared to the open-loop communication setup, by 17 and 23 dB, respectively, a result which is (almost) independent of the codewords length. Also, optimal power allocation increases the diversity gain of the ARQ protocols considerably.
\end{abstract}


%
\IEEEpeerreviewmaketitle
\vspace{-0mm}
\section{Introduction}
Due to the fast growth of wireless networks and of data-intensive applications, green communication and improving the power efficiency are becoming increasingly important for wireless communication. As reported by \cite{Gartner}, the network data volume is expected to increase by a factor of $2$ every year, associated with a $16-20\%$ increase of energy consumption, which contributes about $2\%$ of global $CO_2$ emissions. Hence, minimizing the power consumption is a very important design consideration, and power-efficient data transmission schemes must be taken into account in wireless networks \cite{1321221,5783982,greenref3,greenref4,greenref5}.

From another perspective, automatic repeat request (ARQ) is a well-established approach aiming towards reliable wireless communication \cite{1661837,throughputdef,mimoarqkhodemun,5336856,noisyARQkhodemun,outageHARQ,ARQGlarsson,greenkhodemun,4356994,powerarq2007}. Utilizing both forward error correction and error detection, ARQ techniques reduce the data outage probability by retransmitting the data which has experienced \emph{bad} channel conditions. Consequently, as we show in the following, the joint implementation of adaptive power controllers and ARQ protocols can improve the power efficiency of outage-limited communication systems.

Adaptive power allocation in ARQ protocols is addressed in various papers, e.g., \cite{5336856,noisyARQkhodemun,outageHARQ,ARQGlarsson,4356994,greenkhodemun,powerarq2007}, where the results are obtained under the assumption of asymptotically long codewords. On the other hand, in many applications, such as vehicle to vehicle and vehicle to infrastructure communications for traffic efficiency/safety or real-time video processing for augmented reality, the codewords are required to be short (in the order of $\sim 100$ bits) \cite{wirelessletterkhodemun,4657278,metisdel}. Thus, it is interesting to investigate the performance of power-adaptive ARQ protocols in the presence of finite-length codewords.

In this paper, we study the power efficiency of the ARQ protocols utilizing codewords of finite length. We use the recent results of \cite{5452208,6620483,yanginft} on the achievable rates of finite block-length codes to investigate the power-limited outage probability of the ARQ protocols. The performance analysis is presented for Type-I ARQ where both the error-detecting and the forward error correction information are added to each message and the receiver disregards the previous messages, if received in error.

We investigate the effect of the codeword length on the optimal power allocation and the outage probability of the ARQ protocols. In particular, we show that, for codewords of length $L\ge 50$ channel uses, the performance of ARQ protocols is (almost) insensitive to the length of the codewords, in the sense that the changes in the outage probability are negligible for different codeword lengths. As demonstrated, considerable power efficiency improvement is achieved by the implementation of power-adaptive ARQ. For instance, consider Rayleigh fading channels, codewords of rate $1$ nats-per-channel-use (npcu) and target outage probability $10^{-3}.$ Then, compared to the open-loop communication setup, implementation of ARQ with a maximum of 2 and 3 transmissions reduces the average power by $17$ and $23$ dB, respectively, a result which is (almost) independent of the codewords length. With a maximum of $M=2$ transmissions, we derive closed-form solutions for the optimal, in terms of power-limited outage probability, power allocation between the ARQ transmissions. Finally, it is shown that with a maximum of $M=2$ transmissions the diversity gain of the ARQ protocol increases from 2 to 3, if optimal power allocation is utilized.

\vspace{-0mm}
\section{System model}
Consider a communication setup where the power-limited input message $X$ multiplied by the fading coefficient $h$ is summed with an independent and identically distributed (iid) complex Gaussian noise $Z \sim \mathcal{CN}(0,1)$ resulting in the output
\vspace{-0mm}
\begin{equation}
\vspace{-0mm}
Y = h X + Z.
\vspace{-0mm}
\end{equation}
We study the block-fading conditions where the channel coefficients remain constant in a fading block, determined by the channel coherence time, and then change to other values according to the fading probability density function (pdf). Let us define $g\doteq|h|^2$ which is referred to as the channel gain in the following. The results are given for Rayleigh fading channels where $h \sim \mathcal{CN}(0,1)$ and, as a result, $f_g(x)=e^{-x}$ with $f_g$ denoting the channel gain pdf. In each block, the channel coefficient is assumed to be known by the receiver, which is an acceptable assumption in block-fading channels \cite{1661837,throughputdef,mimoarqkhodemun,5336856,noisyARQkhodemun,outageHARQ,ARQGlarsson,4356994,greenkhodemun,powerarq2007,5452208,6620483,yanginft}. However, there is no instantaneous channel state information available at the transmitter except the ARQ feedback bits\footnote{The transmitter is assumed to know the long-term channel statistics, as it is required for parameter optimization.}.

We consider Type-I ARQ with a maximum of $M-1$ retransmissions, i.e., the data is transmitted a maximum of $M$ times, and in each round the receiver disregards the previous messages, if received in error. Also, we define a packet as the transmission of a codeword along with all its possible retransmissions. Finally, the results are obtained for a frequency-hopping based scheme where the fading coefficient changes in each transmission independently.

\vspace{-0mm}
\section{Performance analysis}
Using power-adaptive ARQ, $b$ information nats are encoded into a codeword of length $L$ channel uses. Thus, the codeword rate is $R=\frac{b}{L}$ npcu. In the $m$th, $m=1,\ldots,M,$ transmission round, the codeword is scaled to have power $P_m$ which, as the noise variance is set to $1$, represents the transmission signal-to-noise ratio (SNR) as well (in dB, the SNR is given by, e.g., $10\log_{10}(P_m)$). The codewords are transmitted until the receiver correctly decodes the data or the maximum permitted transmission round is reached.

If the data is correctly decoded at the end of the $m$th round, the total consumed energy is $\xi_{(m)}=L\sum_{i=1}^m{P_i}.$ Also, the total consumed energy is $\xi_{(M)}=L\sum_{i=1}^M{P_i}$ if an outage occurs, where all possible transmissions are used. In this way, with some manipulations, the expected energy consumed within a packet period is found as
\begin{align}\label{eq:finiteexpectedE}
\bar \xi=L\sum_{m=1}^{M}{P_m\Phi_{m-1}},
\end{align}
where $\Phi_m$ represents the probability that the data is not correctly decoded by the receiver in rounds $n=1,\ldots,m,$ and $\Phi_{0}\doteq1.$

Following the same arguments, the total number of channel uses is $\tau_{(m)}=mL$, if the data transmission is stopped at the end of round $m.$ Hence, the expected number of channel uses within a packet period is given by
\begin{align}\label{eq:finiteexpectedL}
\bar \tau=L\sum_{m=1}^{M}{\Phi_{m-1}},
\end{align}
and the average power, defined in, e.g., \cite{excellentref}, is obtained by
\begin{align}\label{eq:finitepower}
\bar P=\frac{\bar \xi}{\bar \tau}=\frac{\sum_{m=1}^{M}{P_m\Phi_{m-1}}}{\sum_{m=1}^{M}{\Phi_{m-1}}}.
\end{align}
Finally, by the definition, the outage probability is found as $\Pr(\text{Outage})=\Phi_M$ which rephrases the power-limited outage minimization problem as
\begin{align}\label{eq:finiteproblemformulation}
& \mathop {\min }\limits_{ P_m,m=1,\ldots,M} \,\,\,\,\,\,\, \Phi_M, \nonumber\\&
 \,\,\,\,\,\,\,\,\,\,\,\,\,\text{s.t.}\,\,\,\, \,\, \frac{\sum_{m=1}^{M}{P_m\Phi_{m-1}}}{\sum_{m=1}^{M}{\Phi_{m-1}}}=\pi,
\end{align}
with $\pi$ representing the power constraint. As discussed in, e.g., \cite{outageHARQ,ARQGlarsson,greenkhodemun}, (\ref{eq:finiteproblemformulation}) is a complex problem and, depending on the fading pdf and the maximum number of transmissions, there may be no closed-form solution for the optimal powers minimizing the outage probability. Also, note that optimizing the power terms based on (\ref{eq:finiteproblemformulation}) affects the expected delay for a packet transmission and, consequently, the throughput. However, with a limit on the maximum number of transmissions, the expected delay is not of interest in outage-limited data transmission scenario, because the throughput is not an objective function in this case. Moreover, as shown in \cite{greenkhodemun}, unless the SNR is very low, the throughput changes are negligible if, instead of uniform power allocation, the power terms are optimized in terms of power-limited outage probability.

Up to now the results are general in the sense that they are independent of the fading pdf, ARQ protocol and the codewords length. Also, to study the power-limited outage probability of different schemes the final step is to calculate the probabilities $\Phi_m,m=1,\ldots,M.$ For Type-I ARQ, in particular, we have
\begin{align}\label{eq:probphim}
\Phi_m=\begin{cases}
\prod_{j=1}^m{\phi_j}  & \text{ if } m\ne 0 \\
 1 & \text{ if } m=0,
\end{cases}
\end{align}
where $\phi_j$ is the probability that the data is not decoded in round $j$. Here, (\ref{eq:probphim}) is based on the fact that 1) an independent fading realization is experienced in each round, 2) a scaled version of the initial codeword is sent in each transmission of a packet and 3) in each round, the receiver decodes the data only based on the received signal in that round.

In the following, we use the recent results of \cite{5452208,6620483,yanginft} to find $\phi_m$ for the cases with codewords of finite length. Let us first define an $(L,N,P,\epsilon)$ code as the collection of
\begin{itemize}
  \item An encoder $\Gamma:\{1,\ldots,N\}\mapsto\mathcal{C}^L$ which maps the message $n\in\{1,\ldots,N\}$ into a length-$L$ codeword $c_n\in\{c_1,\ldots,c_N\}$ satisfying the power constraint
      \begin{align}\label{eq:code1}
\frac{1}{L}\left \| c_j \right \|^2\le P, \forall j.
\end{align}
  \item  A decoder $\Omega:\mathcal{C}^L\mapsto\{1,\ldots,N\}$ satisfying the maximum error probability constraint
      \begin{align}\label{eq:code1}
\mathop {\max }\limits_{ \forall j}\Pr(\Omega(y)\ne J|J=j)\le \epsilon
\end{align}
with $y$ denoting the channel output induced by the transmitted codeword according to $y.$
\end{itemize}
The maximum achievable rate of the code is defined as
\begin{align}\label{eq:achievablerateeq1}
 R_\text{max}(L,P,\epsilon)=\sup\left\{\frac{\log N}{L}:\exists (L,N,P,\epsilon) \text{code}\right\} \,\,\text{(npcu)}.
\end{align}
Considering block-fading conditions, \cite{yanginft,6620483} have recently presented a very tight approximation for the maximum achievable rate (\ref{eq:achievablerateeq1}) as
\begin{align}\label{eq:achievablerateeq2}
R_\text{max}(L,P,\epsilon)&= \sup\left\{R:\Pr(\log(1+gP)<R)<\epsilon\right\}\nonumber\\&-\mathcal{O}\left(\frac{\log L}{L}\right)\,\,\text{(npcu)},
\end{align}
which, for codes of rate $R$ npcu, leads to the following error probability \cite{yanginft,6620483}
\begin{align}\label{eq:errorfiniteblock}
\epsilon(L,R,P)\approx E\bigg[Q\bigg(\frac{\sqrt{L}\left(\log(1+gP)+\frac{\log L}{2L}-R\right)}{\sqrt{1-\frac{1}{(1+gP)^2}}}\bigg)\bigg].
\end{align}
Here, $U(x)=\mathcal{O}(V(x)),x\to\infty$ is defined as $\lim_{x\to\infty}\sup|\frac{U(x)}{V(x)}|<\infty $ and $E[.]$ represents the expectation with respect to the channel gain $g.$ Also, $Q$ denotes the Gaussian $Q$-function. Note that, according to \cite{yanginft,6620483}, the approximations in (\ref{eq:achievablerateeq2}) and (\ref{eq:errorfiniteblock}) are very tight for sufficiently large values of $L$.

Using (\ref{eq:probphim}) and (\ref{eq:errorfiniteblock}), the probability that the data is not decodable in rounds $n=1,\ldots,m$, i.e., $\Phi_m,$ is found as
\begin{align}\label{eq:probphimfinal}
\Phi_m=\begin{cases}
\prod_{j=1}^m{\epsilon(L,R,P_j)}  & \text{ if } m\ne 0 \\
 1 & \text{ if } m=0,
\end{cases}
\end{align}
from which we can investigate the power-limited outage minimization problem (\ref{eq:finiteproblemformulation}). For instance, using (\ref{eq:finiteproblemformulation}) and (\ref{eq:probphimfinal}), Fig. 1 demonstrates the outage probability of Type-I ARQ with different numbers of transmissions $M$. Here, the results are obtained for codewords of rate $R=1$ npcu and length $L=\{50,200,\infty\}$ channel uses. Also, the optimal power allocation, in terms of (\ref{eq:finiteproblemformulation}), is derived with the same procedure as in \cite[Algorithm 1]{greenkhodemun}. As it can be seen, the system performance is not sensitive to the length of the codewords, for length $L\ge 50$ channel uses. Note that, as the codeword length decreases the tightness of the approximation (\ref{eq:errorfiniteblock}) is reduced. This is the reason why we present the results for the cases with $L\ge 50$ channel uses, for which the approximation is tight enough, and we do not consider shorter codewords. In the meantime, although the approximation is not tight for small $L$'s and the results should not be fully trusted in that case, we observe the same qualitative conclusions as in the case of $L\ge 50,$ when the simulations are run for very short (practically not interesting) codewords (see \cite{6620483,yanginft} for more discussions on the tightness of (\ref{eq:errorfiniteblock}) and \cite{4657278} for practical codes of interest in, e.g., vehicle to vehicle communication).

As demonstrated in Fig. 1, the power efficiency is considerably improved by the implementation of ARQ. For instance, with an outage probability $10^{-3}$ the implementation of ARQ with a maximum of $M=2$ and $3$ transmissions improves the power efficiency, compared to the open-loop setup ($M=1$), by $17$ and $23$ dB, respectively; this is a big step towards green communication. The intuition for the significant performance gain of ARQ is as follows. With an outage probability constraint, the initial transmission(s) of the ARQ is set to have a small power. If the channel is \emph{bad}, the message can not be decoded and is retransmitted. On the other hand, if the channel experiences good conditions, this gambling brings high return. In other words, the ARQ makes it possible to exploit the time diversity and split the total power between the slots which, with high probability, are not used.

\begin{figure}
\vspace{-0mm}
\centering
  \includegraphics[width=1\columnwidth]{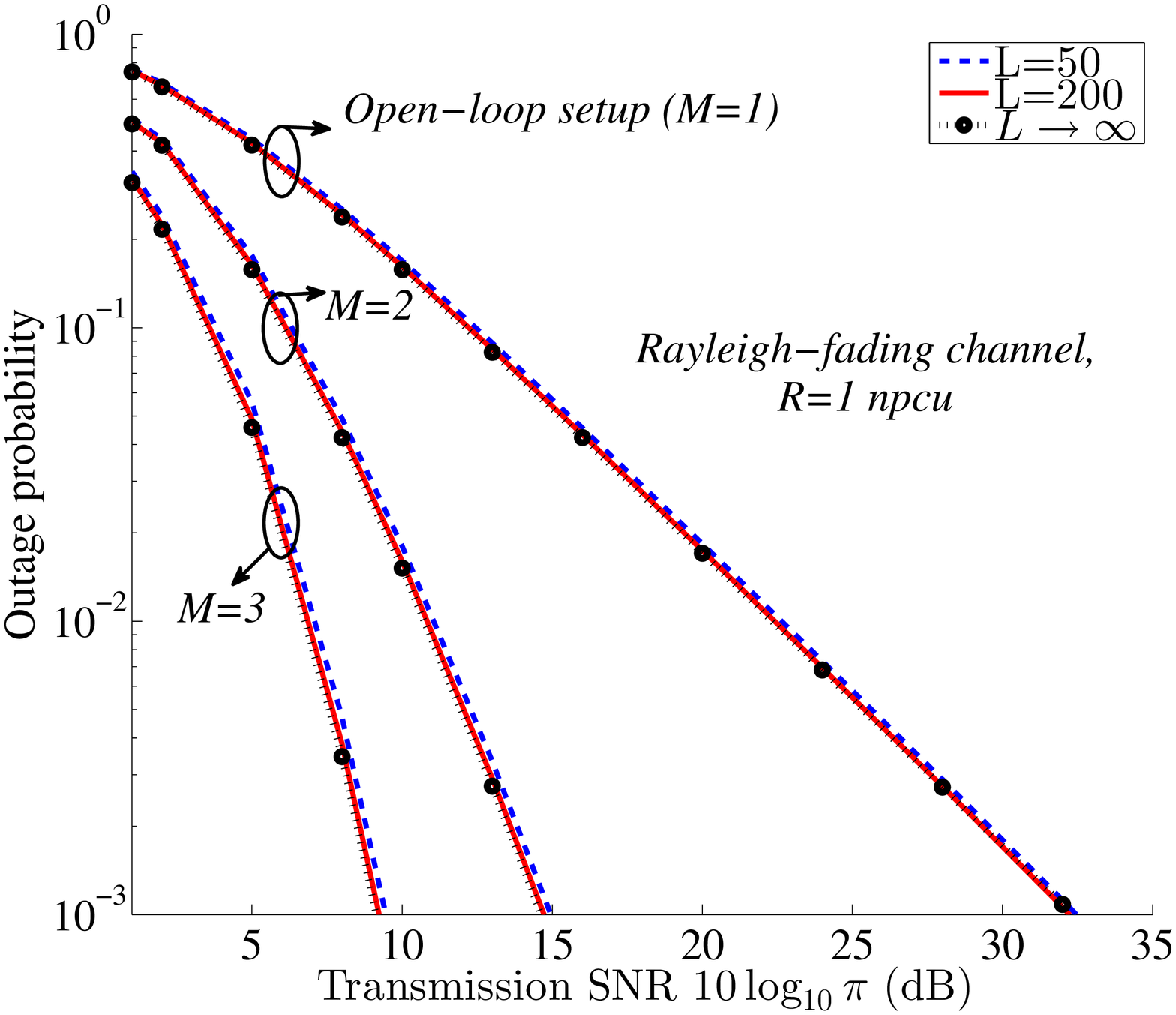}\\\vspace{-0mm}
\caption{Outage probability for different maximum numbers of transmissions. The transmission powers are optimized, in terms of (\ref{eq:finiteproblemformulation}). Rayleigh fading channel, $R=1$ npcu.}\label{figure111}
\vspace{-0mm}
\end{figure}
\subsection{ARQ with a maximum of $M=2$ transmissions}
To further elaborate on (\ref{eq:finiteproblemformulation}), let us concentrate on the case with a maximum of $M=2$ transmissions, for which (\ref{eq:finiteproblemformulation}) is rephrased as
\begin{align}\label{eq:finiteproblemformulationM2}
& \mathop {\min }\limits_{ P_1,P_2} \,\,\,\,\,\,\, \epsilon(L,R,P_1)\epsilon(L,R,P_2), \nonumber\\&
 \,\,\,\,\text{s.t.}\,\,\,\, \,\, \frac{{P_1+P_2\epsilon(L,R,P_1)}}{{1+\epsilon(L,R,P_1)}}=\pi.
\end{align}
Particularly, Theorem 1 studies the optimal power allocation and the diversity gain of Type-I ARQ with a maximum of $M=2$ transmissions. Interestingly, the theorem indicates that, with $M=2,$ the diversity gain of Type-I ARQ is increased from $2$ with uniform (non-adaptive) power allocation to 3, if the powers are optimized in terms of (\ref{eq:finiteproblemformulationM2}).\footnote{Following the same procedure as in Theorem 1 part 3, the diversity gain of the considered ARQ protocol is found as $D=2$ if uniform power allocation, i.e., $P_m=\pi,\forall m,$ is utilized.}

\textbf{Theorem 1.} Considering Type-I ARQ with a maximum of $M=2$ transmissions, the following assertions are valid:
\begin{itemize}
  \item[1)] $P_1\le P_2,\forall \pi,L,R.$
  \item[2)] At high SNRs, the optimal power allocation rule is given by $P_1=\frac{2\pi}{3},$ $P_2=\frac{2\pi^2}{9\theta}, \theta\doteq e^R-1.$
  \item[3)] The diversity gain is $D=3$, if the powers are optimized in terms of (\ref{eq:finiteproblemformulationM2}).
\end{itemize}
\begin{proof}
To prove part 1, we consider two cases, $\{\text{Case 1} : {(P_1 = P + \Delta, P_2 = P)}\}$ and $\{\text{Case 2} : {(P_1 = P, P_2 = P +\Delta)}\}$, $P,\Delta \ge0$, and show that less average transmission power is obtained in the second case. Note that, based on (\ref{eq:finiteproblemformulationM2}), there is no preference between the transmission powers from the outage probability perspective, because the powers are interchangeable in $\Phi_2=\epsilon(L,R,P_1)\epsilon(L,R,P_2)$. Thus, the same outage probability is achieved in the two considered cases. Then, based on the following inequalities
\begin{align}
\pi_\text{Case 1}&=\frac{P+\Delta+P\epsilon(L,R,P+\Delta)}{1+\epsilon(L,R,P+\Delta)}\nonumber\\&\ge \frac{P+(P+\Delta)\epsilon(L,R,P)}{1+\epsilon(L,R,P))}=\pi_\text{Case 2}\nonumber\\&\Leftrightarrow \Big(P+\Delta+P\epsilon(L,R,P+\Delta)\Big)\Big(1+\epsilon(L,R,P)\Big)\nonumber\\&\ge \Big(P+(P+\Delta)\epsilon(L,R,P)\Big)\Big(1+\epsilon(L,R,P+\Delta)\Big) \nonumber\\&\Leftrightarrow 1\ge \epsilon(L,R,P+\Delta)\epsilon(L,R,P),
\end{align}
less average power is achieved in the second case. Therefore, in the optimal case, we have $P_1\le P_2, \forall \pi,L,R.$

Part 2 follows from the fact that at high SNRs the maximum achievable rate (\ref{eq:achievablerateeq2}) converges to the one with asymptotically long codewords, i.e.,
\begin{align}
R_\text{max}(L,P,\epsilon)= \sup\left\{R:\Pr(\log(1+gP)<R)<\epsilon\right\}\nonumber
\end{align}
for $P\to\infty$. Thus, defining $\theta\doteq e^R-1,$ we have
\begin{align}
\phi_m=\Pr(\log(1+gP_m)<R)=1-e^{-\frac{\theta}{P_m}}\nonumber
\end{align}
at high SNRs, where the last equality is for Rayleigh fading channels. In this way, using the Taylor expansion $e^{-x}=1-x,$ the high-SNR outage probability is found as $\Phi_2=\frac{\theta^2}{P_1P_2}$ and the power-limited outage minimization problem (\ref{eq:finiteproblemformulationM2}) is rephrased as
\begin{align}\label{eq:equiproblem}
\left\{ \begin{array}{l}
 \mathop {\min }\limits_{ P_1,P_2} \,\, \frac{\theta^2}{P_1P_2}, \\
 \text{s.t.}\,\,\,\,  \frac{P_1+\frac{P_2\theta}{P_1}}{1+\frac{\theta}{P_1}}= \pi \\
 \end{array} \right.\equiv\left\{ \begin{array}{l}
 \mathop {\max }\limits_{ P_1,P_2} \,\, P_1P_2, \\
 \text{s.t.}\,\,\,\,  P_2= \frac{\pi(P_1+\theta)-P_1^2}{\theta}. \\
 \end{array} \right.
\end{align}
Hence, the optimal power allocation rule is obtained by  $\frac{\partial ((\pi(P_1+\theta)P_1-P_1^3)}{\partial P_1}=0$ which, ignoring its lowest term at high SNRs, results in $P_1=\frac{2\pi}{3},$ $P_2=\frac{2\pi^2}{9\theta}.$

Finally, part 3 is a consequence of part 2; replacing the optimal power terms $P_1=\frac{2\pi}{3},$ $P_2=\frac{2\pi^2}{9\theta}$ into the high-SNR outage probability $\Phi_2=\frac{\theta^2}{P_1P_2}$, the diversity gain $D=-\lim_{\pi\to \infty}{\frac{\log(\Pr(\text{Outage}))}{\log \pi}}$ \cite[eq. 14]{1661837} is found as
\begin{align}\label{eq:diversityARQTypeI}
D=-\lim_{\pi\to \infty}{\frac{\log(\frac{27\theta^3}{4\pi^3})}{\log \pi}}=3,
\end{align}
as stated in the theorem.
 \end{proof}
Here, we should mention that the same result as in part 1 has been previously reported by \cite{powerarq2007} for the cases with asymptotically long codewords. Also, \cite{greenkhodemun} has shown the same conclusion as in Theorem 1 part 1 in the cases with infinitely long codewords and Types-II and -III hybrid ARQ (HARQ). Finally, the theorem emphasizes that the outage probability and the optimal power allocation rule become independent of the codeword length as the SNR increases.

Setting $R=1$ npcu, Figs. 2-4 analyze the performance of ARQ protocols with a maximum of $M=2$ transmissions. Compared to uniform power allocation, i.e., $P_m=\pi,\forall m,$ the optimal power allocation leads to considerable outage probability reduction, especially at high SNRs. Also, setting $L=50$ channel uses, Fig. 3 shows the optimal powers, in terms of (\ref{eq:finiteproblemformulationM2}), and compares the results with those achieved via the theoretical approximations of Theorem 1 part 2. For moderate/high SNR, the approximations match the exact values with very high accuracy. Moreover, the figure emphasizes the validity of Theorem 1 part 1 where $P_1\le P_2,\forall \pi$ (see the black dashed lines in Fig. 3). Finally, Fig. 4 investigates the effect of the codeword length on the outage probability and the optimal power terms of the ARQ protocol.  In harmony with Figs. 1-3, the results emphasize that the system performance is not affected by the length of the codewords, if $L\ge 50$ channel uses.

\begin{figure}
\vspace{-0mm}
\centering
  \includegraphics[width=1\columnwidth]{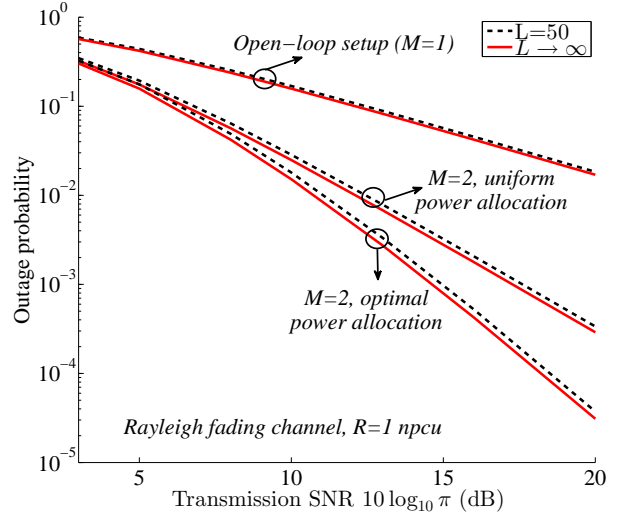}\\\vspace{-0mm}
\caption{Outage probability vs the transmission SNR $10\log_{10}\pi$ dB. Rayleigh fading channel, $R=1$ npcu.}\label{figure111}
\vspace{-0mm}
\end{figure}

\begin{figure}
\vspace{-0mm}
\centering
  \includegraphics[width=1\columnwidth]{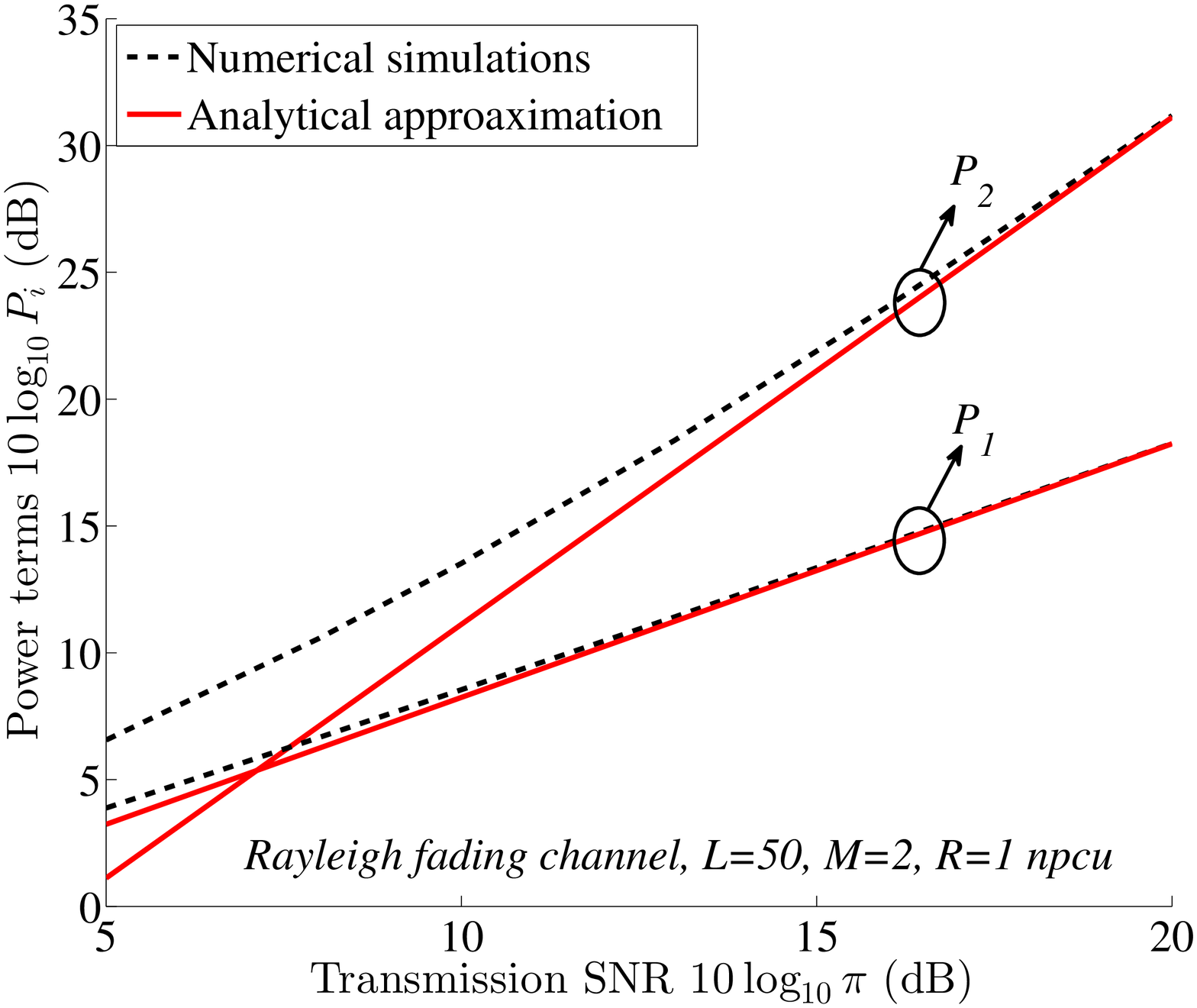}\\\vspace{-0mm}
\caption{Comparison between the optimal power terms derived from numerical simulations and the theoretical approximation of Theorem 1. Rayleigh fading channel, $R=1$ npcu.}\label{figure111}
\vspace{-0mm}
\end{figure}

\begin{figure}
\vspace{-0mm}
\centering
  \includegraphics[width=1\columnwidth]{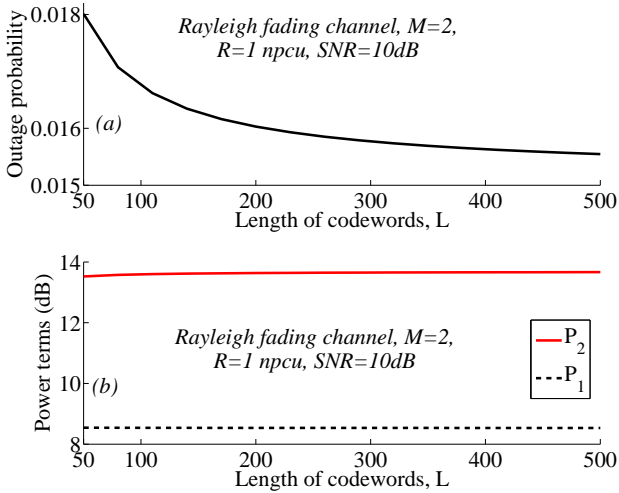}\\\vspace{-0mm}
\caption{(a): Outage probability and (b): the optimal power terms vs the length of the codewords $L$, Rayleigh fading channel, $R=1$ npcu, $\text{SNR}=10\text{dB}.$  }\label{figure111}
\vspace{-0mm}
\end{figure}

\section{Conclusion}
This paper studied the outage-limited power efficiency of ARQ-based systems in the presence of finite-length codes. We utilized the recent results on the achievable rates of finite-length codes to investigate the effect of the codeword length on the performance of ARQ protocols. We showed that, for codewords of length $L\ge 50$ channel uses, the performance of ARQ protocols is (almost) insensitive to the length of the codewords, in the sense that the changes in outage probability are negligible for different codeword lengths. Also, the results show that substantial power efficiency improvement is obtained via the combination of optimal power control and ARQ protocols. The diversity gain of ARQ-based systems is also increased if the power terms are optimally allocated between the transmissions.
\section*{Acknowledgement}
This work was supported in part by the Swedish Governmental Agency for Innovation Systems (VINNOVA) within the VINN Excellence Center Chase.
\vspace{-0mm}
\bibliographystyle{IEEEtran} 
\bibliography{masterfiniteblock}
\vfill
\end{document}